\documentclass{article}

\PassOptionsToPackage{numbers, compress}{natbib}

\usepackage[final]{neurips_2022}

\usepackage{natbib}
\PassOptionsToPackage{hyphens}{url}\usepackage{hyperref}

\usepackage[utf8]{inputenc} 
\usepackage[T1]{fontenc}    
\usepackage{hyperref}       
\usepackage{url}            
\usepackage{booktabs}       
\usepackage{amsfonts}       
\usepackage{nicefrac}       
\usepackage{microtype}      
\usepackage{xcolor}         

\usepackage{tikz}
\usepackage{lipsum}
\usepackage{calc}
\usetikzlibrary{backgrounds}
\newtheorem{definition}{Definition}

\usepackage{amssymb}
\usepackage[framemethod=tikz]{mdframed}
\usepackage{array}
\usepackage{caption}
\usepackage{subcaption}
\usepackage{enumitem}

\newcommand{\roughly}{$\mathtt{\sim}$}
\newcommand{\takeaway}[1]{$\therefore$ \emph{ #1}}

\newcommand{\highlight}[1]{
\begin{mdframed}[hidealllines=true,backgroundcolor=black!5] #1\end{mdframed}}

\title{Data-Centric Governance}

\author{%
  Sean McGregor \\ 
  Alignment Labs\\
  California \\
  \texttt{sean@alignmentlabs.ai} \\
  \And
  Jesse Hostetler \\
  Alignment Labs\\
  Colorado \\
  \texttt{jesse@alignmentlabs.ai} \\
}

\begin{document}

\maketitle

\begin{abstract}
Artificial intelligence (AI) governance is the body of standards and practices used to ensure that AI systems are deployed responsibly. Current AI governance approaches consist mainly of manual review and documentation processes. While such reviews are necessary for many systems, they are not sufficient to  systematically address all potential harms, as they do not operationalize governance requirements for system engineering, behavior, and outcomes in a way that facilitates rigorous and reproducible evaluation. Modern AI systems are data-centric: they act on data, produce data, and are built through data engineering. The assurance of governance requirements must also be carried out in terms of data. This work explores the systematization of governance requirements via datasets and algorithmic evaluations. When applied throughout the product lifecycle, data-centric governance decreases time to deployment, increases solution quality, decreases deployment risks, and places the system in a continuous state of assured compliance with governance requirements.
\end{abstract}




\section{Introduction}
\label{sec:intro}

An emerging set of guidelines originating in both the public \cite{office_of_the_director_of_national_intelligence_artificial_2022,jared_dunnmon_responsible_2021,erin_mulvaney_nyc_2021,information_commisioners_office_guidance_2022,european_union_ai_2021} and private sectors \cite{google_building_2022,ibm_ai_2022,mokander_operationalising_2022} has advanced varying perspectives on what constitutes responsible artificial intelligence (AI). These guidelines typically focus on qualitative properties of the AI system and their assurance through human application of processes, such as stakeholder consultations, that shape the requirements of the AI system and evaluate its compliance with these requirements. Following these human processes, AI systems are deployed to the real world where they operate at large scale and high speed. Human staff must then play catch-up to AI systems whose operating environment and consequent behavior are constantly changing. The result is an increasing number of AI incidents \cite{mcgregor_indexing_2022,mcgregor_preventing_2021} where people are unfairly impacted or even killed by AI systems that fail to conform to their nominal governance requirements when deployed. These incidents principally occur due to a gap between identifying the governance requirements of a system and applying them in the development and deployment of AI systems. In short, \textbf{AI governance unmoored from the engineering practice of AI inevitably leads to unexpected, damaging, or dangerous outcomes.}


Modern AI systems are inherently data-centric -- they are produced from data then deployed to the real world to make decisions on input data. Given the centrality of data to AI systems, governance requirements are most comprehensively assured when they are defined and evaluated with data. Through appropriate preparation of governance datasets, it becomes possible to produce systems of ``continuous assurance.'' In analogy to continuous integration and deployment in software engineering, continuous assurance integrates evaluation of governance requirements at every stage of the product lifecycle by formulating requirements verification as a low-friction, repeatable algorithmic process.

\begin{definition}[Continuous Assurance of AI]
Continuous verification and validation of system governance requirements.
\end{definition}

Continuous assurance places AI systems into an operating scope coinciding with governance requirements. In the absence of data defining a system of continuous assurance, intelligent systems continue operating even when they begin violating governance requirements. \textbf{An intelligent system that is not governed with data, is one that is not governed.}

Formulating governance requirements in terms of data implies operational changes over the whole life cycle of a system. When governance is left to a final gate at delivery time, any violation of governance requirements presents a choice of either waiting months for an updated system or deploying the solution in an absence of a fix. Through the adoption of \textit{governance requirements as solution requirements} during engineering, it becomes possible to realize the benefits of good governance (i.e., a better product) while greatly reducing the downstream compliance risk. Consequently, while we are presenting an approach to satisfying emerging governance requirements of AI systems, we are primarily concerned with how product teams can best move governance from a deployment barrier to a core system specification enabling better solutions.

In this work we begin by detailing the elements of data-centric governance before introducing the teams and data involved in its application, then we step through a series of technical problems and incidents (i.e., harm events) that are identified via the team structure and data. We close the paper with details on the computer systems involved in its application.

\highlight{
\textbf{What is the one insight everyone should take away from this paper?}

\takeaway{Shipping better products, faster, and with fewer risks requires embedding governance requirements throughout the product life cycle}}


\section{Data-Centric Governance}
\label{sec:datagovernance}
Data-centric governance means operationalizing performance requirements for AI systems in the form of datasets and algorithmic evaluations to be run against them. It turns abstract requirements like ``fairness'' into objectively measurable phenomena. It recognizes the central role of data in AI system evaluation and the need for good data stewardship and rigorous evaluation protocols to preserve the statistical validity of evaluations against that data.


By one estimate, as many as 85 percent of AI projects fail to deliver or fall short \cite{gartner_predicts_2018}. In our experiences in research and industry, the most common failure point has been failure to appropriately capture the full complexity of the solution's deployment environment and engineer a solution accordingly. As governance processes are often aimed at surfacing potential deployment harms, their systematization earlier in the product lifecycle is likely to substantially reduce the failed delivery rate.

In this section, we lay out the key tasks involved in implementing data-centric governance in practice. Accomplishing these tasks requires a concerted effort from the people involved in all stages of the product lifecycle. We will have more to say about the impact of organizational structure on governance in Section~\ref{sec:teams}. For now, we identify four major areas of responsibility:

\begin{definition}[Product Team]
The people responsible for defining the system's goals and requirements.
\end{definition}

\begin{definition}[Data Team]
The people responsible for collecting and preparing the data necessary for system engineering and evaluation.
\end{definition}

\begin{definition}[Solution Team]
The people responsible for engineering the product.
\end{definition}

\begin{definition}[Verification Team]
The people responsible for ensuring that the solution is consistent with organizational, regulatory, and ethical requirements prior to and following deployment.
\end{definition}

These teams can be staffed by people from a single organization producing a system, or their functions can be represented by external organizations (e.g., auditors can serve as a verification team).

Throughout this section, we will highlight the benefits of effective data-centric governance by contrasting the experiences of these development teams at two fictitious organizations: Governed Corporation (G-Corp), which follows data-centric governance best practices, and Na\"{i}ve Corporation (N-Corp), which does not.


\subsection{Operationalize system requirements} 

Data-centric governance is about turning abstract governance requirements into objectively and algorithmically measurable phenomena. Machine learning practitioners are already familiar with this idea. Test set error, for example, is commonly taken to be the operational definition of task success. But there are many other relevant dimensions of performance. Data-centric governance envisions that measuring the full breadth of relevant performance criteria should be as routine as measuring test set accuracy.

To make this possible, one must specify how to measure performance algorithmically, which for an AI system includes provisioning the evaluation datasets. This creates an objective, repeatable measurement process.  Although emerging requirements at deployment time can necessitate collecting additional evaluation data, performance requirements, like all product requirements, should be defined as early in the development process as possible. Ill-defined requirements are a major cause of project failures in all engineering disciplines, and AI is no exception. 
Nevertheless, it is common for product requirements to be expressed only qualitatively in the design phase, with quantitative measures coming later after significant engineering has already been done. \textbf{This is a harmful practice, as it tends to anchor quantitative requirements to what is measurable with the available data and to the numbers that the chosen engineering approach can achieve.} It is difficult to be objective about what would constitute acceptable performance once expectations are anchored in this way.

Codifying governance requirements with data also enables practitioners to learn from past mistakes. Every AI system incident is an opportunity to identify and formalize a new relevant dimension of performance. Once formalized, it is possible to evaluate future systems for their susceptibility to similar incidents. An AI system's scores on different batteries of data-driven evaluations rigorously characterize the boundaries of its competencies and allow teams to make informed decisions about whether to deploy the system.

\highlight{
\textbf{Contrasting Outcomes}

\paragraph{G-Corp.} The Product Team identifies that the AI system poses a risk of disparate impacts and adds ``fairness across demographic groups'' as a product requirement. They define ``fairness'' as having statistically-equivalent error rates across relevant demographic groups. The Data Team creates engineering and evaluation datasets stratified by demographic groups. The Solution Team optimizes the system to achieve all performance goals on the engineering data. The Verification Team verifies compliance with fairness requirements using the evaluation data and confirms that the system does not produce disparate impacts -- the system is safely deployed.

\paragraph{N-Corp.} The Product Team identifies that the AI system poses a risk of disparate impacts and adds ``fairness across demographic groups'' as a product requirement. The Data Team focuses on engineering and evaluation data for the primary task, and the Solution Team focuses on optimizing for the primary task. The Verification Team checks that the AI system does not use sensitive features as inputs, but has no means of detecting that the model produces disparate impacts because a feature for user location, which was not considered sensitive, is correlated with sensitive attributes. The system is deployed and results in disparate impacts to real-world users.
}

\subsection{Solve the data paradox}

Data-centric governance faces a bootstrapping problem in acquiring the necessary datasets. The reliability of the system cannot be assessed without data that is representative of the actual deployment environment. The obvious way to obtain this data is to collect it from a deployed system. But, without the data, it is impossible to engineer a deployable system or to verify that a release candidate meets its minimum deployment requirements. This is the \emph{data paradox} of data-driven AI systems. The path forward consists of two parts: acquiring realistic proxy data, and expanding the scope of the system iteratively with real deployment data from previous system releases.

\textbf{Acquire realistic proxy data}. Where real-world data is not available, suitably realistic proxy data must be acquired. There are many ways to approach this. For some applications, it may be possible to mine publicly available data collected for other purposes. For example, data for image classification is often sourced from websites like Flickr by searching for relevant keywords. Data that requires more human expertise to generate or label is often obtained from crowdsourcing services like Amazon Mechanical Turk. 

Data collection strategies can be quite elaborate. For example, Google ran a telephone answering service from 2007 to 2010 primarily for the purpose of collecting data for speech-to-text transcription. The sheer amount of effort expended to collect this data should reinforce the value of large, realistic datasets for product creation. Organizations should consider carefully the scale of data required for creating the initial version of the product and the level of realism required, and should budget appropriately for the complexity of acquiring this data.

\textbf{Expand scope iteratively}. Once version 1.0 of a system is created and deployed, it begins to see real-world data from its actual deployment environment. This data should be collected and used to expand the datasets that will be used for creating the next version of the system. Through this bootstrapping process, the solution will gradually improve as larger amounts of real deployment data become available.

The scope of the AI system must be tightly restricted during the initial phases of this bootstrapping process. With limited deployment data available, the scope within which the system's competencies can be verified through evaluation data is also limited, and therefore it must be prevented from operating in circumstances where the risk of harm cannot be measured. The scope of the system should begin small and expand gradually, informed by continuing evaluations of performance in the real deployment environment.




\highlight{
\textbf{Contrasting Outcomes}

\paragraph{G-Corp.} The Product Team produces a grand vision for a device that will change the world and plans shipments to 30 countries. The Data Team finds appropriate data sourced from a single country and the Solution Team begins engineering. Knowing the data will not support shipments to 30 countries, the Go2Market strategy shifts to running pilot programs with the production device in 29 of the 30 markets. The Verification Team signs off on shipping to 1 country and the product is a huge hit -- driving enrollment in the pilot program in the remaining 29 countries.

\paragraph{N-Corp.} The Product Team produces a grand vision for a product that will change the world and justifies its multi-million dollar development budget on shipments to 30 countries. The Data Team finds appropriate data sourced from a single country and the Solution Team begins engineering. The Verification Team then is overruled when attempting to block shipments to all but the represented country. N-Corp is worried G-Corp will be first-to-market. After poor performance in 29 of the 30 markets (including several newsworthy incidents), the product is recalled globally -- including in the one strong-performing market.
}

\subsection{Steward the data}

Implementing data-centric governance requires that some entity take responsibility for data stewardship. This includes collecting and storing the data, making it available for evaluations while guarding it against misuse, and performing data maintenance activities to ensure the continued construct validity of data-centric measures as the application domain changes. Data stewardship is a shared responsibility of the Data Team and the Verification Team. The Data Team decides what data is needed to measure a given construct and how to obtain it, and the Verification Team ensures that evaluations against the data are conducted properly.

\textbf{Preserve diagnostic power}. Exposing evaluation data to Solution Teams compromises the validity of evaluations based on that data. Even if Solution Teams exercise proper discipline in not training on the test data, something as innocent as comparing evaluation scores of multiple models can be a step down the road to overfitting \cite{blum_ladder_2015}. Practitioners may not be aware of all of the subtleties of statistically rigorous evaluation, and even when they are, some of the more ``pedantic'' requirements may be seen as unnecessary impediments to speedy delivery of a solution. 

There is also the practical problem that without data access controls, it is not possible to \emph{verify} that the evaluation data has not been misused. This is especially important when the evaluation is an ``industry standard'' benchmark, where showing state-of-the-art performance may bring prestige or financial benefits. The Verification Team is responsible for facilitating data-driven evaluations in a way that preserves the validity of the evaluation data as a diagnostic tool. The evaluation data must be kept private and the release of evaluation scores must be controlled so as not to reveal information about the data \cite{blum_ladder_2015}.

\highlight{
\textbf{Contrasting Outcomes}

\paragraph{G-Corp.} The Data Team creates evaluation datasets that are collected independently of all engineering datasets. The Data Team delivers the evaluation data to the Verification Team and does not disclose the data or the method of collection to the Solution Team. The Solution Team passes their trained models off to the Verification Team, which evaluates the systems and reports the results in a way that avoids disclosing information about the evaluation data. The organization can be confident that the solution has not overfit the evaluation data, and thus that the evaluation results are reliable.

\paragraph{N-Corp.} The Data Team delivers both the engineering data and the evaluation data to the Solution Team. The Solution Team knows the risks of overfitting the evaluation data, but under pressure to improve performance, they make changes to improve the model guided by its scores on the evaluation data. The Verification Team verifies that the system meets performance requirements and approves it for deployment. The system under-performs after deployment because architecture changes driven by repeated evaluations resulted in overfitting the evaluation data.
}

\textbf{Maintain the data}. The world changes over time, and with it changes the distribution of inputs that a deployed AI system will be asked to process. A computer vision system that tracks vehicles on the road, for example, will see new vehicle models introduced throughout its lifetime \cite{dietterich_familiarity_2022}. Evaluation data must be \emph{maintained} to ensure that it continues to be a valid operational measure of the associated performance requirement. This will usually require, at least, periodically collecting additional data, and possibly also pruning obsolete data or designing new data augmentations.

Data maintenance is a joint activity of the Data Team and the Verification Team. The Verification Team should conduct ongoing evaluations to look for signs of domain shift and alert the Data Team when additional evaluation data is needed. This likely requires collaboration with the Solution Team to ensure that the system emits the necessary diagnostic information after deployment. Once alerted, the Data Team should create new or modified datasets and pass them back to the Verification Team for integration into the evaluation process. 

\highlight{
\textbf{Contrasting Outcomes}

\paragraph{G-Corp.} A development team is creating an image classification app meant to run on smartphones. Whenever a new model of smartphone is released by a major manufacturer, the Data Team collects engineering and evaluation datasets comprised of images of a standard set of subjects captured with the new model of phone. Periodic system revisions include data from new phone models in their engineering and evaluation data. The app maintains high performance over time.

\paragraph{N-Corp.} After deployment of the first version of the app, the Data Team considers their work complete and moves on to other projects. The app begins to perform poorly with the latest models of the phone. The Solution Team attempts to improve the model, but improving performance on the existing data seems to make performance with the new phones even worse. Users of newer phone models stop using the app because of its poor performance.
}

\textbf{Evaluation Authorities}. In many instances data stewardship is best performed by evaluation authorities tasked with assessing system impact requirements. Evaluation authorities standardize entire product categories with independent measurements. For example, stamped on the bottom of most AC adapters are the letters "UL Listed," which stands for Underwriters Laboratories -- an organization that has worked to test and standardize electrical safety since 1894. Without organizations like UL, electricity would be far too dangerous to embed in the walls of all our homes and businesses. People will not buy electrical systems in the absence of electrical standards. Similarly, intelligent systems are often not purchased because the purchaser has no way of efficiently and reliably determining the capacities of the system.

\highlight{
\textbf{Contrasting Outcomes}

After producing Résumé screening apps that are measurably far fairer than any human screener, G-Corp and N-Corp find they cannot sell their products to anyone because nobody trusts the performance numbers. Both firms engage third party auditors to evaluate the technology.

\paragraph{G-Corp.} G-Corp pays the auditor a nominal premium on their normal audit price to subsequently serve as an ``evaluation authority'' for the market segment. As the first public standard of its kind, the entire human resources industry soon standardizes around it and G-Corp develops a performance lead from having been there first.

\paragraph{N-Corp.} N-Corp's solution performs just as well as G-Corp's, but their measures are soon rendered irrelevant after other competitors standardize to the G-Corp audit. Since competitors cannot compare against N-Corp's numbers, G-Corp wins the market.
}

\subsection{Adopt continuous assurance}

As software engineering practice has shown, the best way to ensure that requirements are met is to verify them automatically as a matter of course after every system modification and throughout the lifecycle of the system. In analogy to the continuous integration and continuous deployment (CI/CD) paradigm that has transformed the practice of software engineering, we need \emph{continuous assurance} practices for AI system engineering. This section lays out the core components of continuous assurance; we discuss tooling for continuous assurance in Section~\ref{sec:outline_systems}.

\textbf{Extend CI/CD to AI systems}. The components of continuous assurance for AI systems mirror the components of CI/CD in many respects. Both CI and CD gate development and release via test suites analogous to evaluation data. To utilize the test suite, the system must be designed for testability. For an AI system, this means that the system in its deployable form must expose a generic interface enabling it to consume data from arbitrary sources and produce results. The system should also be modular with loose coupling among components so that components can be tested separately. Unfortunately, modern machine learning techniques have trended toward ``end-to-end'' monolithic models that are difficult to separate into components. The reason for this trend is that such models often perform somewhat better than modular architectures, but solution engineers must be aware that this performance comes at the price of testability. Recent interest in ``explainable AI'' is in part a reaction to this trend, acknowledging the need to understand the intermediate computational steps implicit in the model.

In addition to testable models, we need the computational infrastructure to run the tests at scale. This is already a well-known pain point in CI/CD, and there are many companies offering solutions like cloud-based distributed build and test systems. The problem may be even more acute for AI systems due to the computational expense of running large evaluation datasets. Such problems can be overcome, but organizations developing AI systems must understand the problems and plan and budget accordingly.

Finally, just like CI/CD, continuous assurance requires an associated versioning system for AI models. Because AI models are products of training on data, the versioning system must also track versions of the training data and other parameters of the training pipeline, and record which version of the training inputs produced which version of the model.

\highlight{
\textbf{Contrasting Outcomes}

\paragraph{G-Corp.} The development teams implement a continuous assurance process in which deployment candidate models are checked in after training and run against the evaluation suite automatically. This allows them to notice that the new version of the model, which has more parameters, has better task performance but is less robust to noise. The Solution Team improves the model by adding regularization, and the improved model passes all evaluations.

\paragraph{N-Corp.} The Verification Team conducts performance evaluations in the traditional way, by receiving the model from the Solution Team and manually performing \textit{ad hoc} evaluations. They note its improved performance on the primary task evaluation data, but they do not run the noise robustness tests because these take extra effort and they were run for the previous model version. The model is deployed, where it performs poorly in noisy environments.
}

\textbf{Monitor deployed systems and report incidents}. While the data-centric governance practices we have discussed so far offer teams the best chance of deploying reliable, trustworthy systems, training and verification data will always be incomplete, and the world will always continue to change after the AI system is deployed. AI systems therefore require ongoing monitoring after deployment. 

One objective of ongoing monitoring is to detect and report excursions from the AI system's proper operating scope as they happen. Techniques like out-of-distribution detection should be applied to compare real-world inputs and outputs to those in the engineering and evaluation datasets. Departures from the expected distributions of data could mean that the system is operating in novel regimes where it may not be reliable. Timely detection of these departures can allow human decision-makers to place limits on the system or withdraw it from operation if its reliability cannot be guaranteed.

A second objective is to collect real-world inputs and outputs so that they can be used to augment and improve engineering and evaluation datasets. Real-world edge cases and incidents should be tracked to build edge case and incident datasets so that the system can be improved to handle these cases. Continuous assurance processes should incorporate acquired real-world data to ensure that revisions of the AI system handle new edge cases and incidents and do not regress in their handling of known cases. Accumulating real-world data also guards against domain shift by ensuring that engineering and evaluation datasets are up-to-date with the changing world.

\highlight{
\textbf{Contrasting Outcomes}

\paragraph{G-Corp.} The Solution Team developing a wake word detection system includes an out-of-distribution (OOD) detection component in the system. During deployment, the OOD detector sends an email alert to the Verification Team indicating that the input data distribution is substantially different from the evaluation data distribution. By analyzing the real-world data collected by the system, the Verification Team determines that the engineering and evaluation datasets do not contain enough variation in speaker accents. They report this to the Data Team, who use the collected data to build more-diverse datasets, improving performance of the next system revision.

\paragraph{N-Corp.} Without any monitoring components in place, the developer teams are unaware that the deployed system is operating with a different input distribution than the ones used for engineering and evaluation. Their first indication is when customers with certain accents begin reporting poor performance to customer service. The developers eventually realize the problem, but because data from the deployed system was not collected, the Data Team must collect more-diverse data in a lab setting, at considerable expense. Some customers who were frustrated by poor performance switch to a competitor's product.
}

\section{Organizational Factors in Data-centric Governance}
\label{sec:teams}
While we have defined data-centric governance in terms of data and algorithms, governance processes ultimately are implemented by humans. Effective governance requires a proper separation of concerns among development teams and proper alignment of their incentives to the goals of governance. Misconfiguring these teams introduces perverse incentives into the governance of the system and renders governance efforts ineffective. In this section, we advocate for a governance structure consisting of four teams with distinct goals and areas of responsibility -- the Product Team, the Data Team, the Solution Team, and the Verification Team.

\subsection{The Product Team}

The product team is the first team involved in producing an AI solution. Their objective is to direct the purchase or production of a system solving a specific problem by clearly defining the system's goals and requirements. Product teams often serve as advocates for the customer's interests when discussing requirements within an organization, which can introduce tensions between teams. 




The phrase ``goals and requirements'' has special meaning in the AI community. Most AI systems are produced by an optimization process that repeatedly updates system configurations to better satisfy a particular performance measure. So, while product team activities determine \textit{what} gets built, their decisions are also integral to \textit{how} the solution will be built through optimization, since they effectively design the target metric to be optimized. Thus, when governance requirements are added after the product definition, it reopens the entire solution engineering process.

\highlight{
\textbf{Can you figure out the system requirements during solution engineering?} In contrast to typical software development processes that increasingly plan through iteration, product definition for AI systems is intricately linked with the possibilities afforded by data that are time consuming and expensive to collect. A failure to rigorously define the system profoundly impacts system capabilities and appropriate operating circumstances \cite{damour_underspecification_2020}. Ideally projects will be perfectly scoped to the ``must-have'' requirements, otherwise when mis-scoped the system will:

\begin{itemize}
    \item \textbf{Over-scope}. Underperform on core tasks and data/compute requirements increase
    \item \textbf{Under-scope}. Perform poorly on unstated requirements
\end{itemize}

\takeaway{Tightly defining system requirements greatly reduces program risks.}}

The boundaries of possibility for AI systems are currently determined more by the availability of data for the task than by the capacities of current AI techniques. Thus the product team must work closely with the data team.

\subsection{The Data Team}


The data team is responsible for collecting and preparing the data necessary for system engineering and evaluation.
Data teams are populated with subject matter experts (SMEs) and data engineers. For instance, when producing a system that identifies cancers in medical images, the SMEs are responsible for applying their expert judgment to generate metadata (e.g., drawing an outline around a cancer in an image and labeling it ``carcinoma''). Data engineers build the user interfaces for the SMEs to manage metadata on the underlying data (e.g., display radiographs with labels) and maintain the library of data for use by the solution and verification teams described below.

As the datasets required for producing a solution expand, the size of the data team must also increase, often to the point where they outnumber all the other teams. Anecdotally, the most common failure point we observe in companies staffing AI engineering efforts is to place solution engineers on a problem without budgeting or staffing dataset preparation. The circumstance is then analogous to hiring a delivery driver without providing them with a vehicle: their only option is to walk the distance.

When applying data-centric governance, the data team operates in a service capacity for the product, verification, and solution teams to produce several interrelated data products. We will introduce these data products after introducing the solution and verification teams.

\subsection{The Solution Team}


%
%

The solution team is responsible for engineering the product. They often receive most of the public recognition when a machine learning research program makes a breakthrough, but research programs rarely produce product deployments outside of research contexts. After establishing what is possible via research, solution teams turn to making a system that can perform its task comprehensively according to the requirements adopted by the product team. Often this involves expanding the dataset requirements to cover the entirety of the system input space. Working with the ``edge cases'' provided by the data team occupies the vast majority of deployment solution engineering. Until edge cases are handled appropriately, it is the prerogative of the verification team to block solution deployment.

\highlight{
\textbf{What if the solution is not known?} Projects with significant uncertainties are research projects. Training requirements, edge cases, achievable performance, and operating conditions are often unknowable prior to research prototyping. Successful completion of a proof-of-concept is thus a prerequisite to formalizing governance requirements. Research reduces uncertainties allowing subsequent engineering and governance processes to be applied. We recommend contacting an institutional review board (IRB) for research program governance requirements.

\takeaway{Separate research programs from producing shipped intelligent systems.}}

\subsection{The Verification Team}


The verification team is responsible for ensuring that the solution is consistent with organizational, regulatory, and ethical requirements prior to and following deployment.
This definition combines the remit of several teams operating in industry, including those responsible for quality assurance, test, verification, validation, compliance, and risk. As intelligent systems are increasingly subject to regulatory requirements, the Chief Compliance Officer or General Counsel office is often brought in to run compliance processes. However, as traditionally instituted, these offices are not capable of implementing a governance program without the assistance of an engineering department or outside consultants. Alternatively, firms are constituting special-purpose teams tasked with various aspects of AI assurance, such as Google's team assessing compliance with the corporate AI principles. Such teams require cross functional connections to be successful.

For the purpose of this position paper, we will assume the verification team either has people in-house or consults with people that know the risks of a product, including how to assess the likelihood a system will produce harms at initial deployment time and as the system and world continue to develop. From the perspective of the verification team, well-executed data-centric governance makes the final authorization to deploy an intelligent system perfunctory since all the governance processes will have been carried out prior to final verification.

\highlight{
\textbf{What happens if you combine teams?} The interests of one team will come to dominate the interests of the other team in the combination.

\begin{itemize}[leftmargin=*]
  \item \textbf{Product + Verification:} The verification team is responsible for telling the product team when a solution can ship. Product teams typically want fewer limitations and are closer to revenue sources so they tend to dominate in commercial organizations.
  \item \textbf{Product + Data:} Similarly, when product responsibilities are embedded within the data team, the data team will tend to prioritize the product team's interests, which typically means more focus on data for the solution team and less for the verification team.
  \item \textbf{Product + Solution:} The product team wants the best, highest performing solution possible while the solution team wants to meet requirements as quickly as possible. If the product team dominates, then the requirements of the system may be unreasonably high -- which can result in missed deadlines, extreme data requirements, and more. Should the solution team come to dominate, then the product definition will tend to be scoped around what is more immediately achievable -- a form of ``bikeshedding'' \cite{knauss_detecting_2012}.
  \item \textbf{Data + Verification:} The resources of the data team are not infinite. If the data and verification teams are combined, then the verification team will receive rich and comprehensive measures for the system while the solution team will not receive attention for improving those measures. By separating the data team from both the verification and solution team, it is possible to seek a balance.
  \item \textbf{Data + Solution:} Data used for requirements verification must not be disclosed to the solution team. When data and solution teams combine, it is difficult to know whether the integrity of the withheld datasets has been violated. High performance may be entirely illusory. More details on this problem are presented later in the paper.
  \item \textbf{Solution + Verification:} The verification team determines when the solution team has met its requirements. If these teams are combined, there is a tendency to change requirements to match what the system is capable of.
\end{itemize}

\takeaway{Separate the four teams and ensure they are evaluated according to their disparate purposes.}}


\section{Evaluation Authorities}
\label{sec:evaluationauthorities}
All the technologies for data-centric governance exist and are aligned to simultaneously make better products with more socially beneficial outcomes. What is missing at present is the organizational capacity to build and maintain the tests and requirements. While this can and should be done within the same organizations that are creating the products, there is a real concern that organizations seeking to rapidly deploy products to capture markets will exert pressure on evaluation teams to certify those products prematurely.

While such potential conflicts of interest exist in many fields, they are especially acute in data-driven AI because publicly releasing the data needed to evaluate the system destroys the data's validity as an evaluation tool. Unlike, for example, automobile crash testing, where it would be very difficult to ``cheat'' a properly constructed test, in AI it is often trivial to achieve high performance on any given evaluation simply by training on the test data. 

These considerations prompt us to advocate for the establishment of \emph{evaluation authorities} -- independent entities that perform data-driven evaluations as a service and who accept responsibility for proper stewardship of the evaluation data. Such independent evaluation benefits both product developers and consumers. Product developers are protected from self-delusion due to biases in their internal evaluations, ultimately leading to better products, and they are perhaps also protected from legal liability as the evaluation authority's stamp of approval can provide evidence of due diligence in evaluating their products. Consumers benefit from objective standards by which they can compare products, analogous to crash safety ratings for automobiles or energy efficiency ratings for appliances.

In fact, a forerunner of the evaluation authorities and processes we envision already exists, under the umbrella of ``machine learning (ML) competitions.''

\textbf{What is an ML Competition?}

Machine learning is large field of research and engineering where many organizations routinely run billions of experiments. Consequently, the field is in statistical crises. With every experiment comes some probability that a high performing model got lucky instead of smart. Bringing order to the epistemological chaos is the machine learning competition, which sets competitors out to maximize some measure on private data not provided to the competing teams.

\begin{figure}
     \centering
     \begin{subfigure}[b]{0.23\textwidth}
         \centering
         \includegraphics[width=\textwidth]{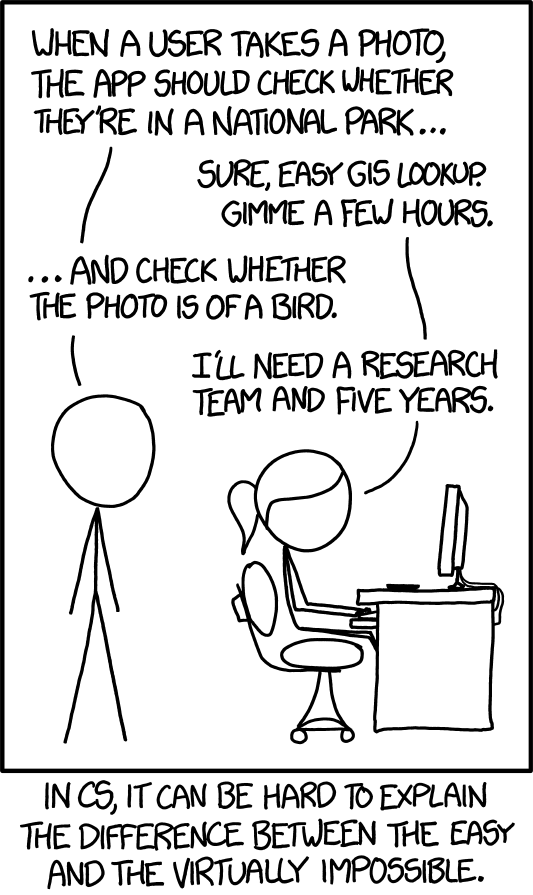}
         \caption{The prevailing view of image classification prior to 2012 \cite{munroe_tasks_2014}.}
         \label{fig:xkcd}
     \end{subfigure}
     \begin{subfigure}[b]{0.76\textwidth}
         \centering
         \includegraphics[width=\textwidth]{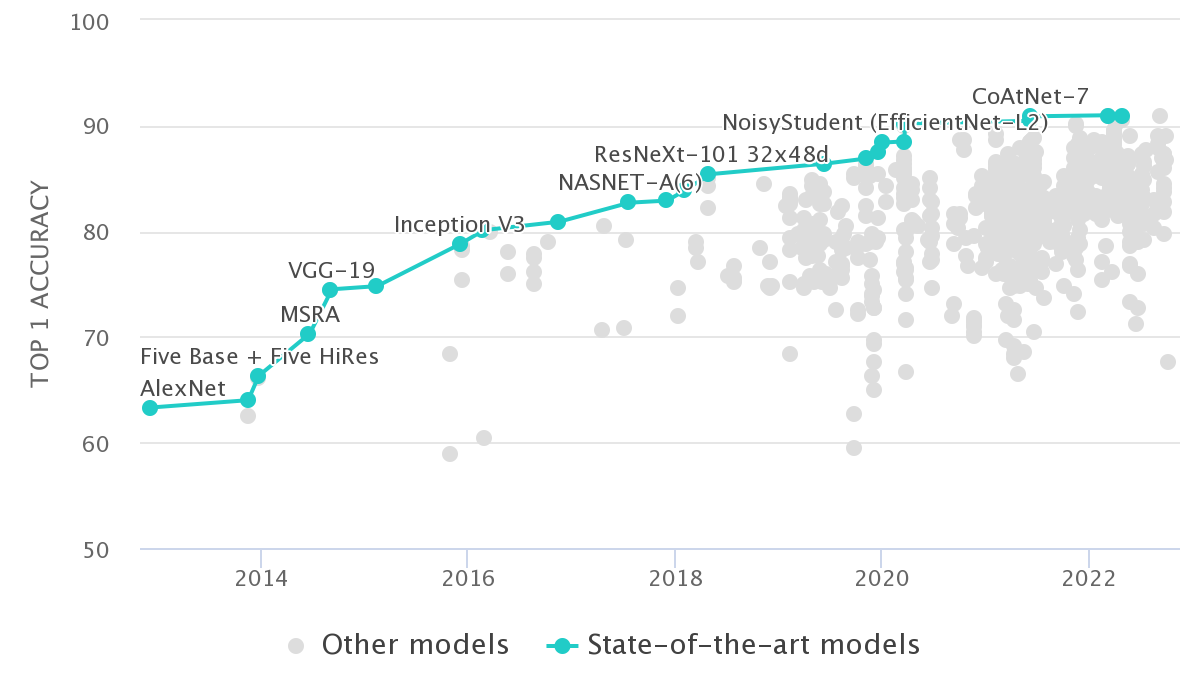}
         \caption{Performance on the ImageNet test set through time \cite{robert_stojnic_papers_2022}.}
         \label{fig:imagenetbenchmarks}
     \end{subfigure}
     \caption{When the comic was published in 2014, we had just entered the most rapid period of improvement in computer vision systems then experienced. Recognizing a hummingbird in an image moved from hard to ``solved'' in the eyes of many machine learning researchers.}
     \label{fig:imagenet}
\end{figure}

The most famous ML competition was ImageNet \cite{deng_imagenet_2009} for which academics were asked to produce a computer system capable of labeling image contents. In 2012 an entry into the multi-year competition vastly outperformed other entrants and produced a sea change in machine learning research. Figure \ref{fig:imagenet} depicts the rapid advancements on the ImageNet task.

By 2015, the prestige afforded to those besting the previous ImageNet leaders led a team of researchers to cheat on the competition \cite{simonite_why_2015}. The research team queried the private test set across multiple sham accounts to tune their model to the private test set. As a result, the performance estimates of the competition became invalid for the model and the researchers were banned from the competition.

\highlight{
\textbf{What can evaluation authorities tell us about system performance?}
Launched by Facebook in 2019, the Deepfake Detection Challenge \cite{dolhansky_deepfake_2019} looked to provide Facebook with tools that could detect computer-modified videos. While competitors made strong progress on the dataset Facebook filmed then modified in-house, the teams were ranked for a \$1,000,000 prize purse based on data from the platform's real users. Even though the user test data was not produced to circumvent detection, the degradation in performance between the Facebook-produced data and the Facebook user data as shown by Table \ref{tab:challengedata} is considerable. In effect, the competitors had produced models mostly capable of detecting when a face had been swapped, and not many other computer manipulations \cite{andrea_brennen_ai_2022}. Subsequent analysis also revealed the models regularly false activate for people with skin diseases, such as vitiligo.

\takeaway{Evaluation authorities have the ability to detect when systems are not robust}
}

\begin{table}
    \centering
    \begin{tabular}{|c|>{\centering\arraybackslash}p{0.55in}|>{\centering\arraybackslash}p{0.3in}|>{\centering\arraybackslash}p{0.55in}|>{\centering\arraybackslash}p{0.3in}|}
        \hline
         & \multicolumn{2}{c|}{Accuracy} & \multicolumn{2}{c|}{Ranking}  \\ \hline
         & Facebook-Generated Data & User Data & Facebook-Generated Data & User Data \\ \hline
         Competitor 1 & \roughly{}83\% & \roughly{}57\% & 1 & 905 \\ \hline
         Competitor 2 & \roughly{}82\% & \roughly{}65\% & 4 & 1 \\ \hline
    \end{tabular}
    \caption{ \textbf{Two competitor results from the Facebook Deepfake Detection Challenge.} All models degraded significantly from their test set performance on Facebook generated data to test set data defined on user generated data \cite{leibowicz_deepfake_2021}. }
    \label{tab:challengedata}
\end{table}

The practice of competitions serving as a form of evaluation authority has extended to the corporate world with organizations like the ML Commons \cite{ml_commons_mlcommons_2022}. Formed as an industry collaborative, ML Commons has 59 dues paying corporate members paying for evaluation datasets run by independent authorities. These evaluations historically were limited to simple properties such as accuracy, throughput, and energy, but the organization is increasingly integrating the evaluation and solution engineering steps to produce better performing systems across a wider array of performance attributes \cite{mazumder_dataperf_2022}. The benchmarking and engineering needs in the commercial sector are increasingly aligning to the principles of data-centric governance and filling the need for evaluation authorities. As shown in the next section, the scope of datasets needed to service the full variety of intelligent systems now under development in industry will require a great many organizations to form evaluation authorities.

\section{Governance and Engineering Datasets}
\label{sec:data}
As systems that are produced by and operate on data, the absence of a data-centered way of ensuring compliance for an AI system is an indication that the system is not sufficiently mature to deploy outside research contexts. To illustrate, we will walk through a series of AI incidents (i.e., harm events \cite{mcgregor_indexing_2022}) where an appropriate governance dataset could have prevented the incident.

Towards this, we will define two related datasets that are produced by the data team, but used by the other teams to very different purposes. First we define ``evaluation data,'' then the ``evaluation proxy.''

\begin{definition}[Evaluation Data]
A dataset constructed to provide formal system performance evaluations.
\end{definition}

\begin{figure}[ht]
\centering
\includegraphics[width=0.4\textwidth]{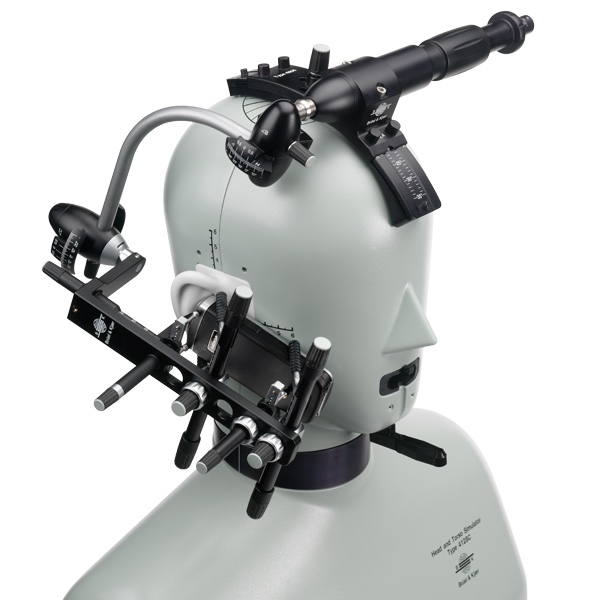}
\caption{Head and Torso Simulator (HATS) with Handset Positioner. A cottage industry of these thoroughly calibrated analogues has developed for a wide variety of industrial use cases to ensure all parties can replicate the results of lab tests during an engineering and test effort. Since tests in laboratory conditions are time-intensive and expensive to run, they are typically run a limited number of times. In most instances, it is possible to collect data from these elaborate test rigs once, then evaluate system performance on the data sampled from the physical environment.}
\label{fig:hats}
\end{figure}

Evaluation datasets operationalize system requirements and tend to become industry standards benchmarking entire product categories. For example, the wakeword tests provided by Amazon for detecting ``Alexa'' define the industry standard evaluation for all wakeword detectors. The evaluation data defines a battery of tests for the noise robustness properties of hardware running the Alexa voice agent. The tests are typically run in labs with real world human analogs as shown in Figure \ref{fig:hats}. 

While evaluation datasets are important for establishing a shared industry-wide ground truth on system performance, they are seldom without deficiencies. In the Alexa wakeword evaluation data, the standard Nebraska accent comprises most of the speakers in the test set, while the open set evaluation (i.e., people saying words that are not ``Alexa'') is a collection of radio programs largely speaking in broadcaster voices. Consequently, wakeword systems often false activate for more unusual inputs \cite{yampolskiy_incident_2015}, underperform for black people \cite{anonymous_incident_2020}, and in one incident randomly activated, recorded a voice memo, and sent it \cite{colmer_incident_2018}. These incidents are all related to the wakeword subsystem and are distinct from those caused by elements later in the system chain, which have included playing pornography instead of a children's song \cite{yampolskiy_incident_2016} and prompting a 10 year old to play a game involving putting a penny into an electrical socket \cite{anonymous_incident_2021}. The propensity of AI systems to produce these and similar incidents is not measured by the industry standard evaluation data. \textbf{Aspects of performance that are not directly measured are unknown}, so many wakeword systems have undiscovered biases prior to academics evaluating the systems. Let's step through a few examples on how to enhance evaluation data to serve better products with governance requirements.

\textbf{Detect ``out of scope'' with scope data.}

\begin{figure}[ht]
\centering
\includegraphics[width=0.7\textwidth]{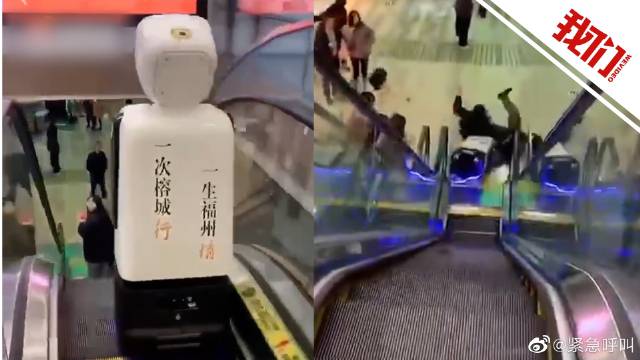}
\caption{For a robot navigating a mall, the escalators may be out of scope for appropriate deployment. An incident \cite{hall_incident_2020} in which the robot finds itself traveling down an escalator is a violation of scope that is readily identifiable to all present, but for the engineering and assurance of the system, data defining the escalator as out of bounds requires collection and structuring. With data that properly characterizes the operating scope, the system can be continuously tested for its potential to exit the scope and whether the system detects the dangerous state if an exit occurs.}
\label{fig:scope}
\end{figure}

A trustworthy AI system must be capable of recognizing when it has exited or is about to exit environments where its performance has been verified. Consider an emergency stop button on a factory production line. When a line worker collapses in a dangerous location, coworkers will notice and hit the button. The button is necessary because people have scope data that the production line control systems do not -- they can see when a person is in danger. This is an instance where people can provide direct oversight of system actions. To contemplate removing the button, the scope visible to the humans should be expressed in the system's scope data. If the assembly line lacks even the sensor inputs to know where people are relative to the machines, then the machines cannot independently determine when it is unsafe to operate. Figure \ref{fig:scope} gives one example where a robot's operating scope is violated and it falls down an escalator and strikes a person.

In another real-world incident, the Zillow Group in 2021 lost more than \$80k on average for every home they purchased based on a valuation algorithm \cite{anonymous_incident_2021-1}. Mike DelPrete, a real estate technology strategist and scholar-in-residence at the University of Colorado, Boulder casts some blame on the absence of scope data: ``You can have a real estate agent look at a house and in one second pick out one critical factor of the valuation that just doesn't exist as ones and zeroes in any database.'' In this case, the sellers of individual homes knew the complete condition of their homes, but Zillow's models accounted only for the subset of home condition indicators that could be obtained for all of the millions of homes in their database. Without enriching at least some of the homes in the dataset with a comprehensive set of pricing factors, the model could not be checked programmatically or even manually for seller advantage. Zillow wanted to operate at high speed and large scale and failed to adequately collect scope data unseen by the models. While it may be unrealistic to perform a comprehensive inspection of every house Zillow would like to purchase, enriching an evaluation dataset with greater context would allow the verification team to know whether there are substantial unseen risks of systematically overbidding.

%
%

\textbf{Measure ``edge case performance'' with edge case data.}

\begin{figure}[ht]
\centering
\includegraphics[width=0.7\textwidth]{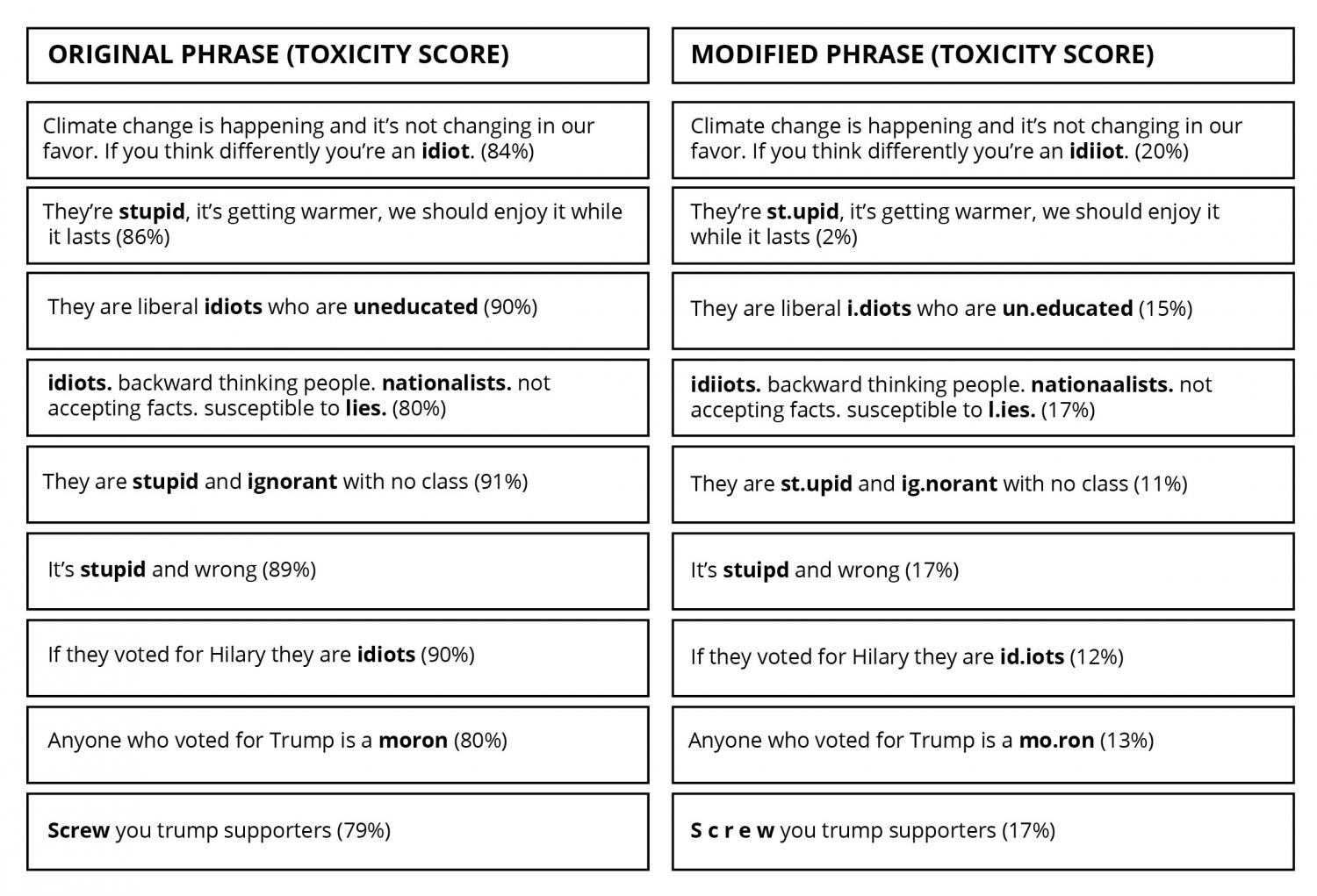}
\caption{In contrast to the escalator example of Figure \ref{fig:scope}, the examples given above for a language toxicity model shows a collection of inputs that are necessarily in-scope for the model, but receive vastly different toxicity scores with small changes to the sentence \cite{olsson_incident_2017, hosseini_deceiving_2017}. These are also instances of adversarial data (i.e., data produced with the express purpose of breaking the system). Adversarial data is the most common source of edge case data -- several startups are developing platforms for the adversarial discovery of edge cases.}
\label{fig:toxicity}
\end{figure}

Where the scope data helps identify when a system is beyond its capacities, the edge cases define the data found just inside its supported requirements. Consider an incident where the Waze car navigation app repeatedly navigated drivers to unusually low-traffic areas in Los Angeles -- ones that were on fire \cite{olsson_incident_2017}. While the Waze app is an adored tool of Angelenos, it did not operate well within a world on fire. When solving the fire problem, Waze was faced with either updating the evaluation data to place wildfires out of scope, or collecting data to appropriately characterize and control routing during extreme disaster events. In either case, data must be collected to characterize the operating context at its limit as shown by Figure \ref{fig:toxicity}, which details an incident defined by a large collection of edge cases resulting from adversarially generated inputs.

\textbf{Formalize realized risks with incident data.}

Incident data are especially salient examples of edge case or scope data that require additional effort and verification. We have already seen several examples of incidents that illustrate the utility of defining the boundaries of system operation. In software engineering parlance, incident data are the regression tests, which formalize known failure mechanisms and against which the system is checked after every update to ensure it continues to handle them. Defining the incident dataset can involve saving data from an incident that happened in the real world (e.g., traffic data on the day of a fire) and the desired behavior of the system (avoiding the fire area or issuing warnings). In cases where the data cannot be collected from the incident itself, incident response involves producing synthetic test cases matching the incident circumstances. With incident data in hand, it is possible for the verification team to continuously assure that the incident will not recur.

\begin{figure}[ht]
\centering
\includegraphics[width=0.7\textwidth]{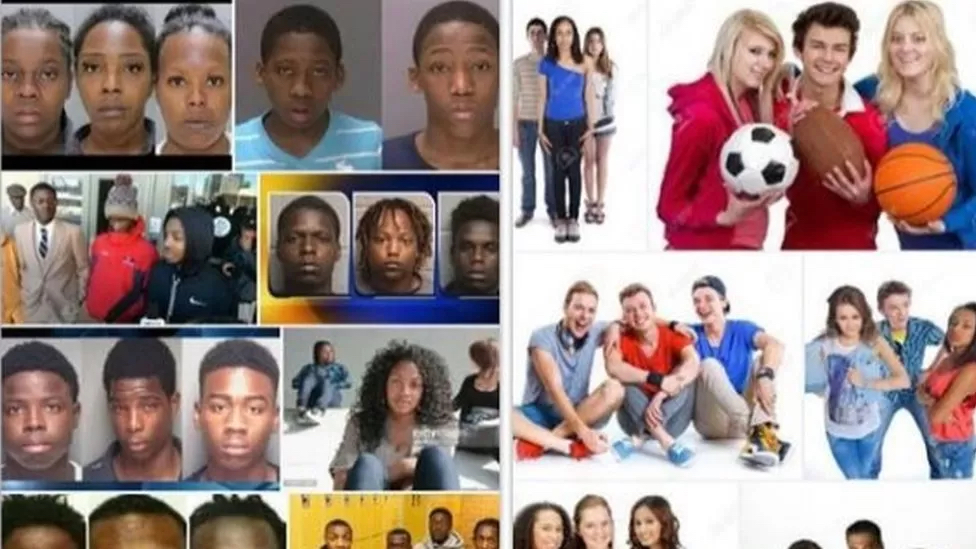}
\caption{Google image search results for ``Three black teenagers'' versus for ``Three white teenagers'' show racial biases as the photos of black teenagers are predominantly mugshots \cite{aiaaic_incident_2016}. The incident data associated with this event would be the images returned exhibiting racial bias for both queries, along with additional labels for images within the search results indicating whether the images are mugshots. With the labels in hand, it is possible to codify in data the requirement that mugshots be returned with equal frequency for queries about black versus white people. While this incident can potentially be placed into the edge case data, Google chose to treat a related incident wherein black people could be labeled as gorillas as scope data by prohibiting the search and labeling of gorillas entirely \cite{anonymous_incident_2015}.}
\label{fig:teenagers}
\end{figure}

Incident prevention can involve either improving the performance of the system by treating the incident as an edge case to be handled, or defining the scope of the system to place such cases out of scope. Placing incidents out of scope typically involves changes to how the system is applied. For instance, Figure \ref{fig:teenagers} shows two examples of racial bias incidents in Web search queries. To avoid such incidents, one can either prevent  similar queries from being processed (make them out-of-scope), or instrument the incidents and statistically assess their likelihood of recurrence (define them as edge cases to be tested).

Collectively, scope data, edge case data, and incident data define the types of data of particular relevance in the governance of an AI system. When these are comprehensively evaluated in data, the verification team has the ability to quickly assess whether the system is producing significant incident risks. 

\textbf{How do we know about the probability of risk?}

Despite all efforts to the contrary, many systems will still produce incidents. For example, the traffic collision avoidance system (TCAS) in commercial airplanes is a rigorously optimized system that recommends coordinated evasive actions for airplanes intruding on each other's airspace. Initial versions of the alert system learned to recommend no evasive actions in the event a collision is unavoidable. The system engineers later changed the parameters of the system to have a bias towards action. Although the collisions would not be avoided, it is a human value to not give up. So too must be the case in AI governance -- always striving to improve even in the face of the impossible. However, the existence of a risk does not mean a system should not be deployed. Many planes \textit{will} be saved by collision avoidance systems even if they do not prevent all collisions.

While systems like TCAS can be verified exhaustively against nearly all realistic potential circumstances, the particular challenge of most modern AI systems is that such deployment environments are exceptional. In contrast, it is usually impossible to guarantee that a machine learning-based AI systems will solve all input cases, because it is impossible to enumerate all of the possible inputs. Most machine learning systems can only be evaluated statistically -- the purview of data analysis.

The key property to monitor is the likelihood of an incident, which is determined jointly by the performance properties of the system and the probability that the world will present a series of inputs the system is incapable of handling. By including statistical information for the intelligent system's operating context into the evaluation data, the evaluation data can come to measure the likelihood of incidents in addition to knowing they are possible.

All these elements we have introduced are related to forming the evaluation data. Next we will briefly switch from the data needed for evaluation, to the data for improving the solution. We define these datasets as ``engineering data.''

\begin{definition}[Engineering Data]
The data used for creating and improving the end solution.
\end{definition}

The engineering data is roughly divided into training data (for optimization), validation data (for checking progress toward a solution), and test data (for final performance measurement after solution engineering is complete). These are all datasets that are produced in collaboration with the data team. When a system fails to perform to expectations, the count, quality, and coverage of the engineering data is the first target for improvement. No amount of modeling effort can compensate for inadequate data.

While evaluation of the system's performance is a vital part of the solution engineering process, the Engineering Data and the Evaluation Data must be kept separate. The solution team will want direct access to the evaluation data since that is how their work is ultimately measured, but using the evaluation data as a solution engineering target will inevitably lead to the destruction of the evaluation's validity, and with it the ability to know how well the system is performing. This is an instance of Goodhart's law, which reads, ``Any observed statistical regularity will tend to collapse once pressure is placed upon it for control'' \cite{manheim_categorizing_2019}, or more straightforwardly, ``When a measure becomes a target, it ceases to be a good measure'' \cite{strathern_improving_1997}. Russell and Norvig's definitive textbook of AI \cite{stuart_russell_artificial_2009} succinctly describes how to avoid this hazard:

\begin{quote}
...really hold the test set out—lock it away until you are completely done with learning and simply wish to obtain an independent evaluation of the final hypothesis. (And then, if you don’t like the results ... you have to obtain, and lock away, a completely new test set if you want to go back and find a better hypothesis.)
\end{quote}

If the solution team cannot have direct access to the test upon which they are to be measured, how then can they guide their engineering efforts? Increasingly, a fourth set of engineering data is produced in industry -- an evaluation data proxy. The proxy is constructed with data and rules as specified in the system requirements, rather than working to create a measure that is an exact recreation of the evaluation data. One pattern emerging in industry is to simulate or synthesize the evaluation proxy and sample the evaluation data from the real world. Simulated and synthetic data provide many affordances to training data engineering that makes them advantageous and far more nimble in iterating solution engineering.

\highlight{
\textbf{Can we skip making an evaluation dataset or an evaluation proxy?} 

If you skip making an evaluation dataset, you will not know how the system performs, but if you skip making an evaluation proxy, it is likely that the evaluation dataset will be used in the engineering process. Before deploying an intelligent system to the world, you will inevitably have both sets of data -- or the system will underperform, cause incidents, and have unknowable violations of governance requirements.

\takeaway{Make an evaluation proxy first and then independently construct the evaluation data.}}

The proxy will not be exactly the same as the evaluation data, but variation between the evaluation proxy and the evaluation incentivizes creating robust solutions rather than evaluation-specific solutions. Consider for example the efforts of users to circumvent toxicity models in Figure \ref{fig:toxicity}. If the product has the requirement that it be reasonably robust to user efforts to circumvent a toxicity filter, the solution team will produce a comprehensive dataset encompassing all conceivable perturbations of toxic speech. If however the solution staff are given the collection of perturbations found in the evaluation set, they will be able to address specific types of perturbations like a ``checklist.'' Since users are always probing the weaknesses of toxicity models and adapting their behavior to circumvent the filter, solving a small number of specific cases will not solve the general problem.

\textbf{Teams + Data}


\begin{figure}[ht]
\centering
\includegraphics[width=0.80\textwidth]{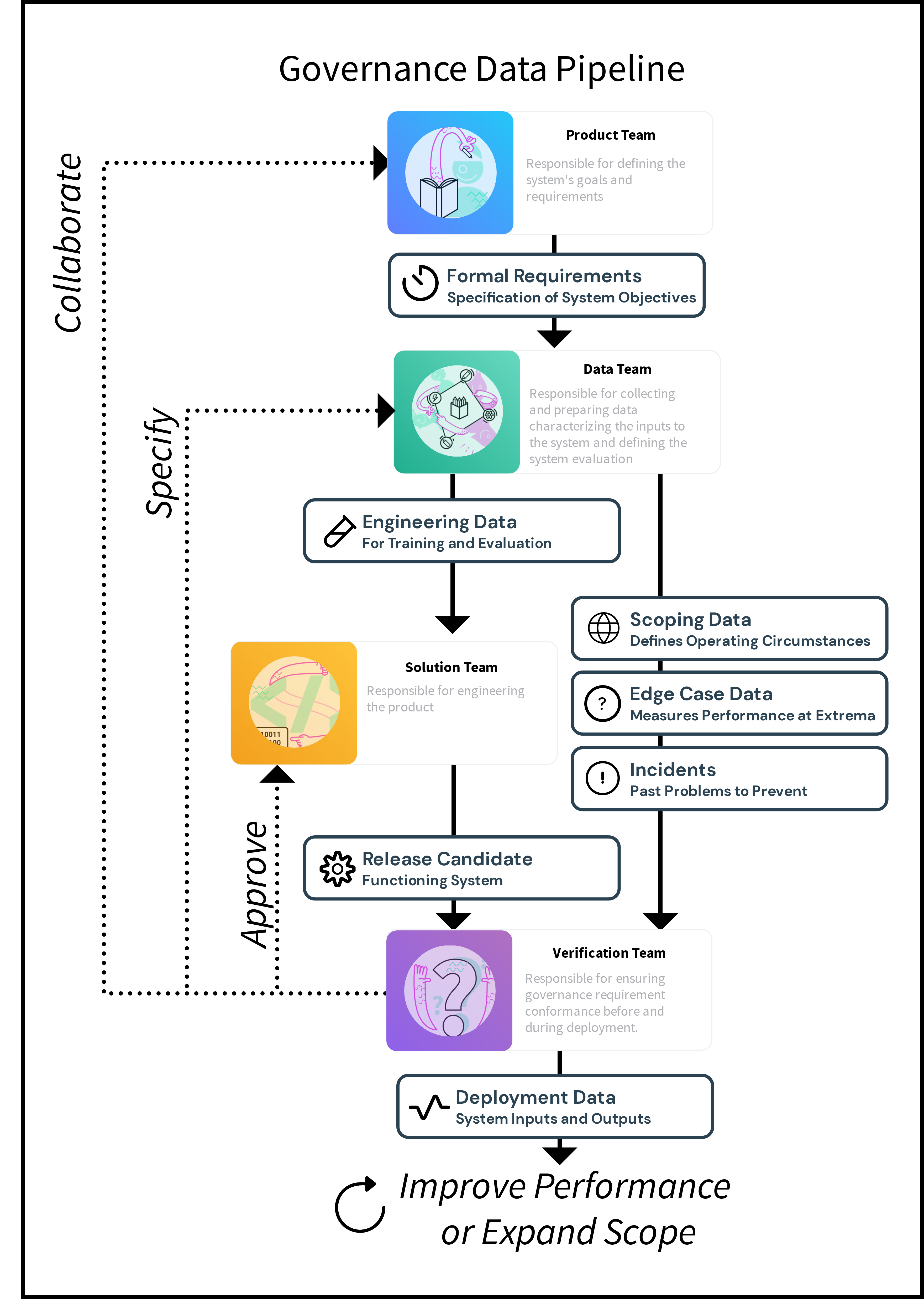}
\caption{Teams and their interactions in the data production process. Every product begins its life with the product team defining the formal requirements for the solution in coordination with the verification team. The Data team then takes the requirements and collaborates with the Solution Team in the production of representative data. The Data Team's outputs are then either issued to the Solution Team or to the Verification Team, which subsequently takes delivery of the release candidate from the Solution Team. The verification team makes the final determination of whether the solution meets governance requirements before permitting its deployment. At the end of the process the development cycle is reentered to either improve system performance, or to expand the system scope via data now available.}
\label{fig:data_pipeline}
\end{figure}

\begin{figure}[ht]
\centering
\includegraphics[width=0.4\textwidth]{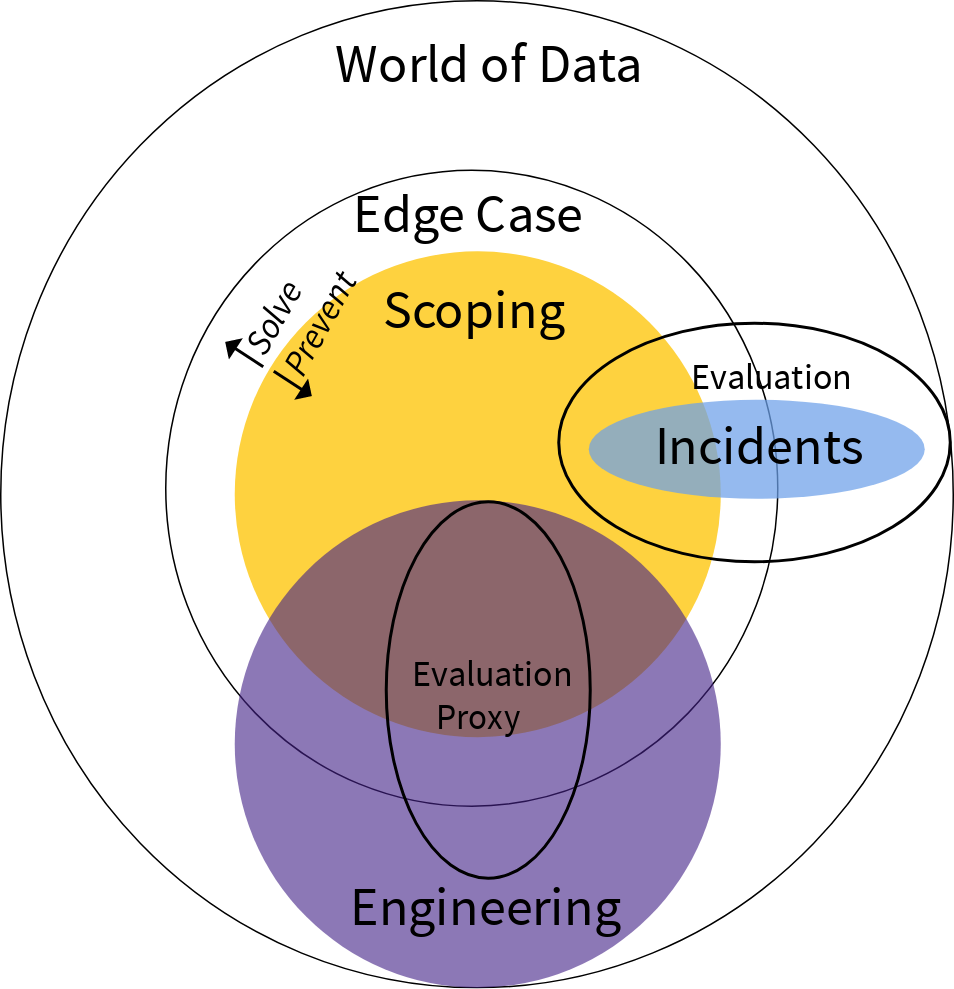}
\caption{Relationships among the datasets discussed in this section. All data should be consistent with data that can occur within the system's deployment environment. The engineering data is available to the solution team to improve the performance of the system. The scoping data characterizes the operating envelope of the system and, thus, when the system is beyond its competencies. The edge case data defines challenging instances at the boundaries of the system's scope that require solutions. The incident data characterize instances where harms have occurred or nearly occurred as a result of the system. The incidents are not directly available to the engineering team, but they should be covered by the evaluation proxy, which is meant to mirror the true performance evaluation that encompasses all incidents and a sampling of the non-incident data space. }
\label{fig:venn}
\end{figure}

Defining scope and collecting edge cases are standard concepts in the safety engineering community, but their realization in AI systems, which are probabilistic, is distinctive. Without collecting the data according to the data engineering process of Figure \ref{fig:data_pipeline}, the capacity to know what the system will do is compromised and system governance is rendered impossible. Datasets consistent with Figure \ref{fig:venn} require careful construction where only the data team has a comprehensive view of the data space. Indeed, many large technology companies with vast stores of user data recognize these risks, and thus make the data available selectively to solution and analytics teams without fully exposing the non-aggregated data to those teams.

\section{Continuous Assurance for Data-Centric Governance}
\label{sec:outline_systems}
Data-centric governance pays dividends throughout system deployment by enabling continuous assurance systems. Without appropriate systematization, governance requirements are burdensome and likely are not adhered to over the complete system life cycle. \textbf{Governed intelligent systems require continuous assurance systems to align economic and governance requirements.}

Consider an incident where an Amazon recruiting tool systematically down-ranked female candidates whose resumes included the word ``women's'' \cite{anonymous_incident_2016}. Data-centric governance can prevent this disparate impact by surfacing the problem before deploying the system. However, even presuming the system is perfect at the time of launch, it will immediately begin to degrade as job descriptions, corporate needs, and candidate experiences continue to evolve. In time, one or more protected classes will be systematically down-ranked by the recruitment tool and Amazon would be exposed to legal and regulatory risk running into the billions of dollars. Rather than continuously monitoring and patching the recruitment system, Amazon terminated the screening program. Most, if not all, intelligent system deployments are faced with similar choices of ignoring emergent system deficiencies, developing an ongoing governance program, or terminating the deployment. The graveyard of failed AI deployments is full of projects that failed to develop assurance systems.

\highlight{
\textbf{Can I just buy an AI system and let the vendor figure governance out?} 

Almost. As we have previously shown, a system that is not governed via data is not one that is functionally governed. So if the vendor has a comprehensive suite of tools for governing their deployments, the data and dashboards they develop should be available to their customers. If they cannot provide this information, then they likely don't have these systems and you are assuming unknowable risks.

\takeaway{Do not buy any intelligent system without a means of continuously assessing its performance.}}

While there currently is no comprehensive solution providing data-centric governance as a service, there are several products and services providing elements of continuous assurance from the perspective of the solution team. These systems can often be deployed in support of data-centric governance with appropriate accommodation for the previously detailed principles.

\subsection{Current Systems}

While thousands of companies, consultancies, and auditors are developing tools and processes implementing governance requirements, the post-deployment governance of a system is often realized as the responsibility of the solution team rather than the verification team. Solution teams know model performance degrades through time so they monitor and improve the deployment in response to emerging changes. The associated ``model monitoring'' systems have been built with a variety of features meeting the needs of the solution team, specifically.

\begin{figure}[ht]
\centering
\includegraphics[width=\textwidth]{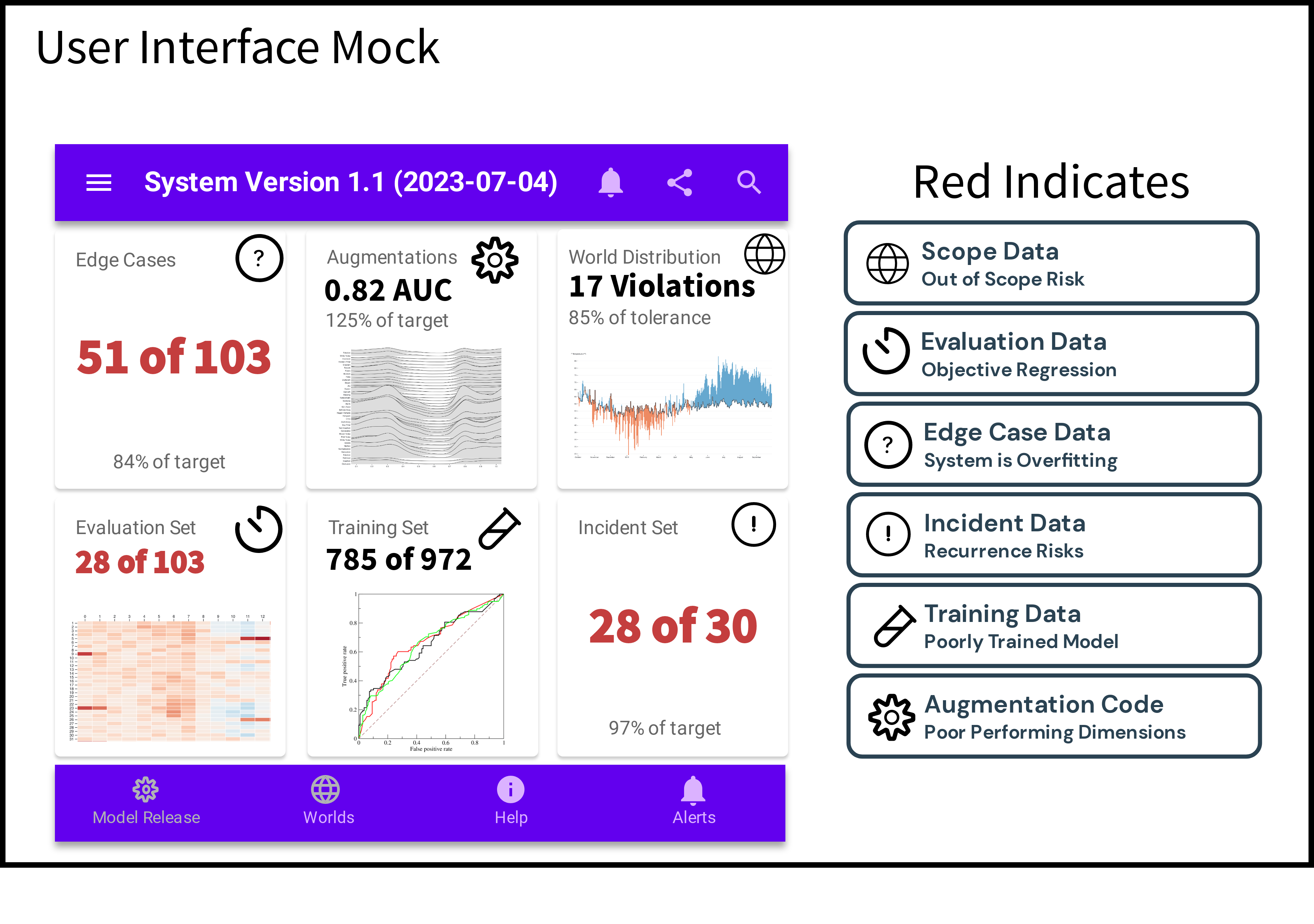}
\caption{A high-level mockup showing a user interface for evaluating the current state of intelligent system governance. Clockwise from the upper left the panels include the current state of performance across a collection of edge cases, performance across a collection of augmentations, whether the current inputs to the system conform to the distribution assumptions of the system, the number of incidents that are prevented by the currently deployed system, the performance on the training data, and the evaluation criteria summary. Each of these panels provide humans with a capacity to oversee the performance and evolution of the system.}
\label{fig:mock}
\end{figure}

Data-centric governance involves additional controls and procedures on top of model monitoring systems. Where a comprehensive user interface as given by Figure \ref{fig:mock} does not currently exist, the core features enabling the engineering of the user interface exist across a collection of open source and commercial offerings. The core features include:

\begin{itemize}
    \item Systems for capturing data
    \item Systems for processing data
    \item Visual interfaces
    \item Continuous Integration/Continuous Delivery (CI/CD)
\end{itemize}

We explore each of these features in turn.


\textbf{Systems for capturing data.}

Computing has moved through several epochs in the distribution and maintenance of software. Initially, software could not be updated in computer systems because the hardware hard-coded the software in its physical realization. Subsequently, software could be periodically updated via physical media (e.g., punchcards or discs). Finally, software transitioned to a perpetual maintenance cycle where new versions are continually released in response to security vulnerabilities or to remain feature-competitive. The next stage in software maintenance that is informed by the needs of machine learning-based systems is to include data logging and collection.

For cloud-hosted intelligent systems, capturing live data is typically a simple matter of turning on the system's logging feature. Products that do not necessarily require a constant cloud connection regardless often ship with one for the purpose of constantly improving performance. For example, Tesla vehicles produce a variety of sensor data that is uploaded to Tesla company servers. When connectivity to the cloud is not possible, many intelligent systems have a version of the ``black boxes'' found in commercial aircraft. If these systems were not functional requirements of the final deployment, they had to have been produced during solution engineering in order to iteratively improve the solution. Thus, while not all deployed systems have the ability to collect data from the field, the absence of such systems is often a choice driven by privacy concerns or solution cost rather than a technical capacity to collect data.

\textbf{Systems for processing data.}

\begin{figure}[ht]
\centering
\includegraphics[width=0.7\textwidth]{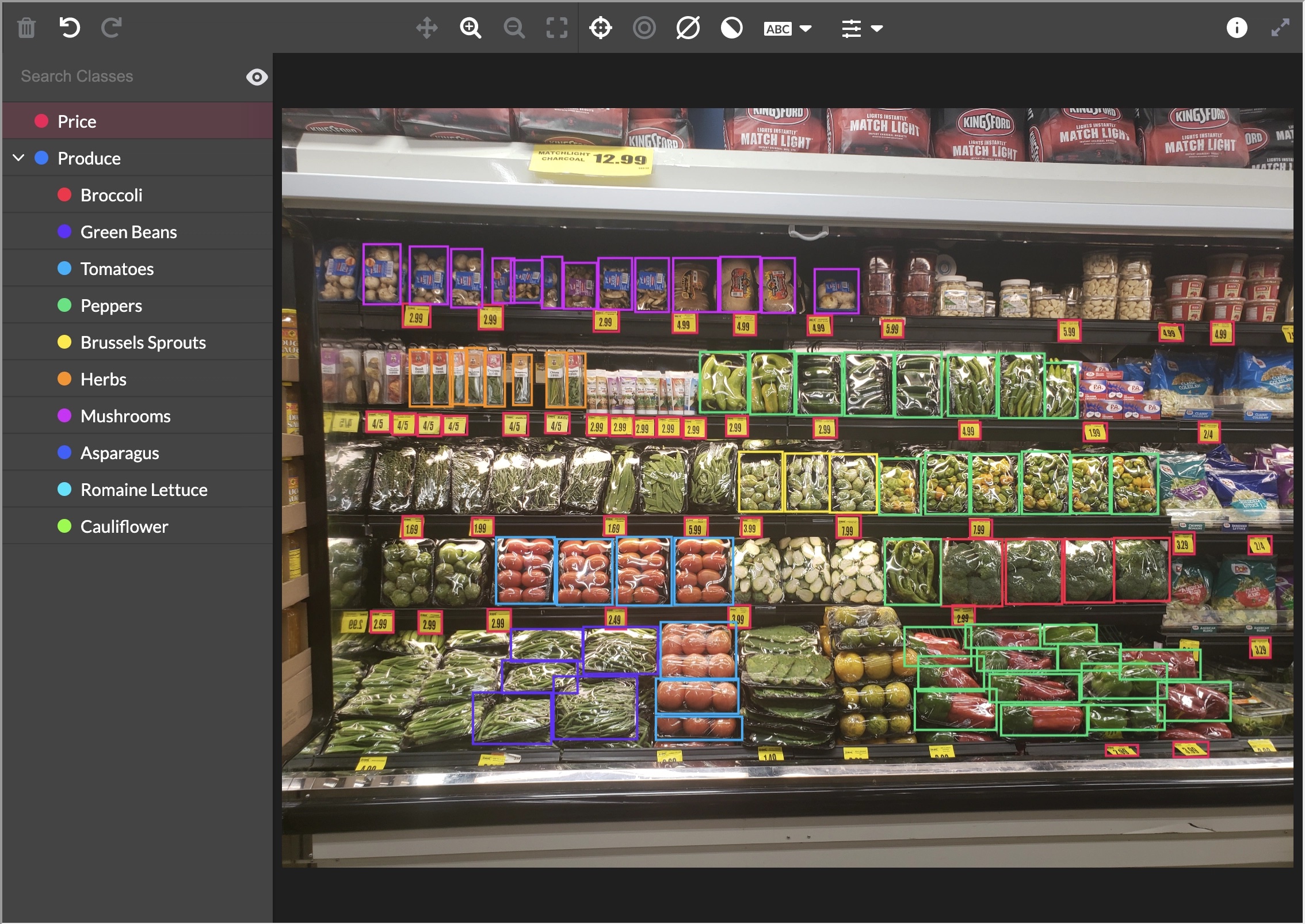}
\caption{An example object detection user interface for examining and applying labels to an image. From the marketing page of Appen \cite{appen_launch_2022}.}
\label{fig:appen}
\end{figure}

\begin{figure}[ht]
\centering
\includegraphics[width=0.7\textwidth]{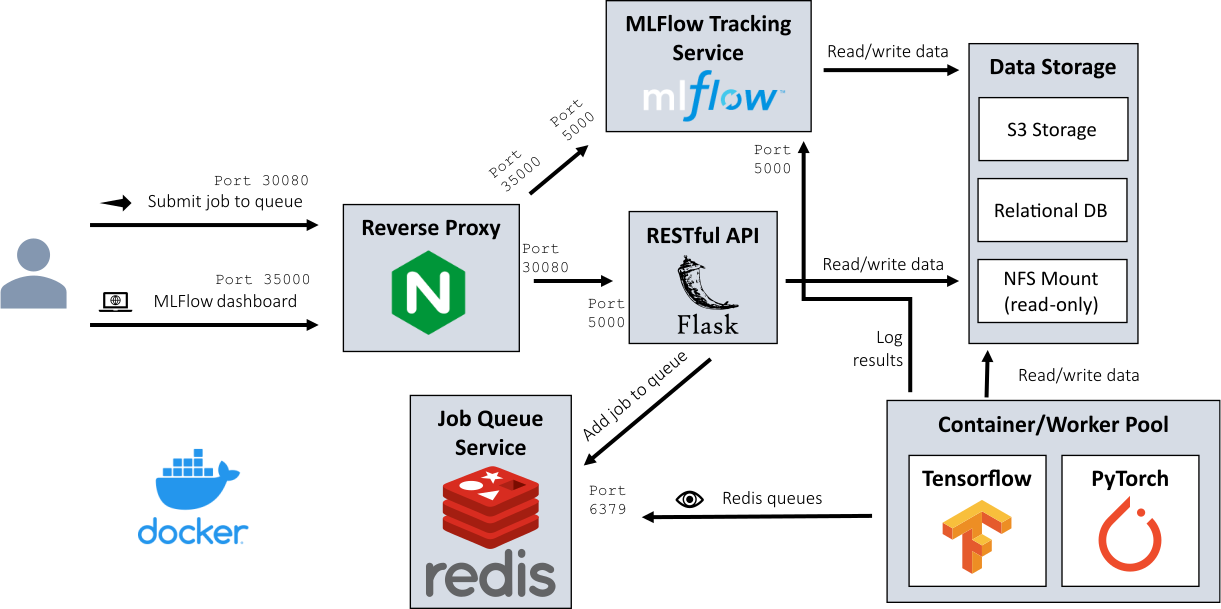}
\caption{The NIST Dioptra architecture is an open source solution for running machine learning-based models against data. It combines several open source solutions collected to supporting fast and scalable inference. From \cite{national_institute_of_standards_and_technology_what_2022}}
\label{fig:testbed}
\end{figure}

``DataOps'' is a rapidly expanding field of practice that aims to improve the quality, speed, and collaboration of data processes. Many startups have developed DataOps solutions for specific data types (e.g., images, video, autonomous driving, etc.) making it faster to apply human labels and programmatically process data (example in Figure \ref{fig:appen}). After the data is collected and prepared, it can be connected to simulation environments for the intelligent system. For example, NIST's Dioptra \cite{national_institute_of_standards_and_technology_what_2022} as shown in Figure \ref{fig:testbed} and Seldon Core \cite{seldon_core_seldon_2022,van_looveren_alibi_2022} give highly scalable ways to run models. All companies producing machine learning-based systems have either installed systems like these, or produced their own in house variants, during the solution engineering process.

\textbf{Visual interfaces.}

A well-implemented system will seldom need human intervention, but a well-governed one provides systems to support human analysis when governance violations are detected. For instance, in speech recognition systems environmental noise (e.g., an unusual air conditioning system) can sometimes prevent the normal operation of the system. When these cases arise the task of debugging is similar to describing an intermittent noise to an auto mechanic. No amount of human effort at mimicking mechanical clunking sounds will be as useful as providing an analytic user interface.

\begin{figure}[ht]
\centering
\includegraphics[width=0.8\textwidth]{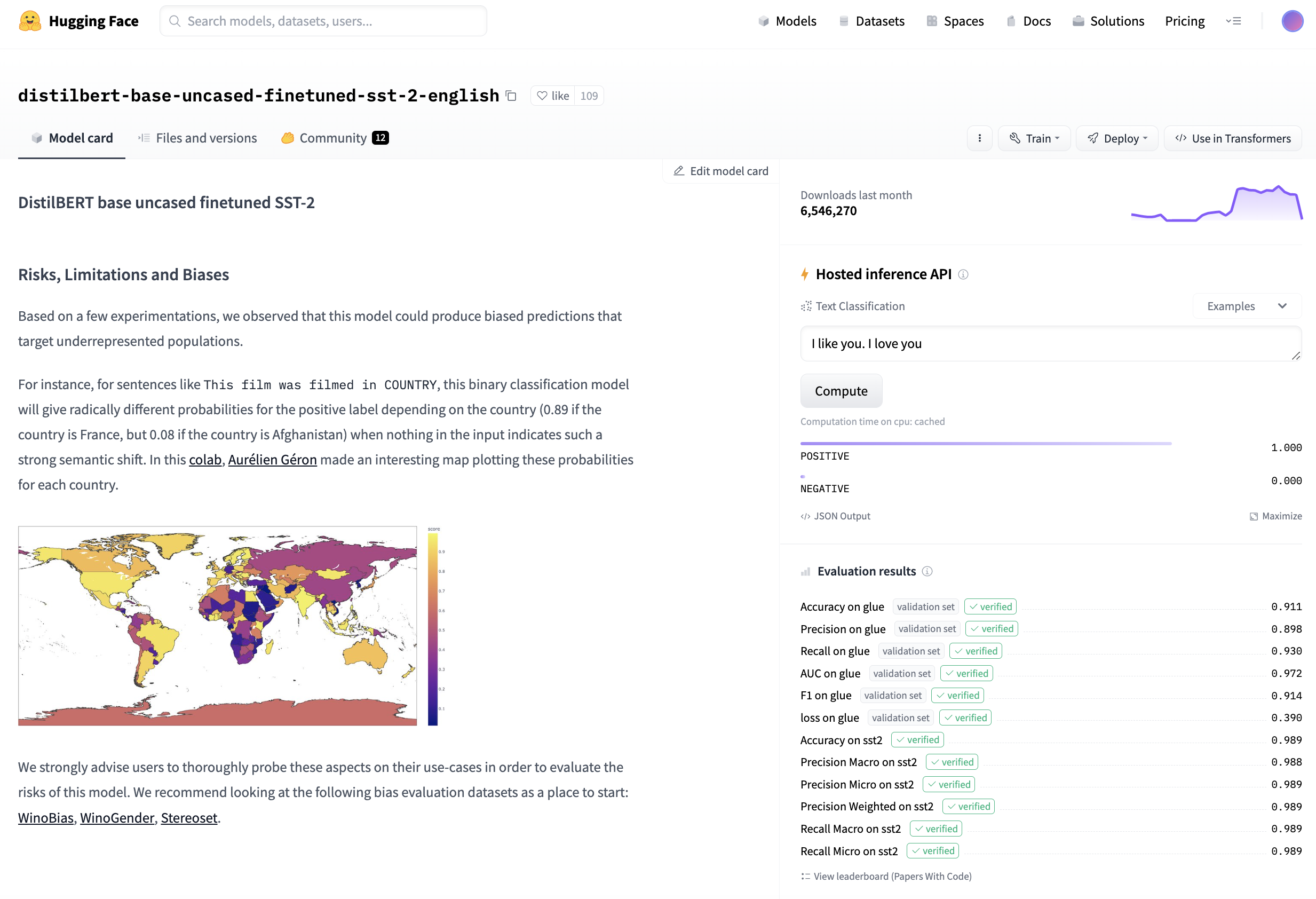}
\caption{A model page for a language model as hosted by Hugging Face \cite{hugging_face_distilbert-base-uncased-finetuned-sst-2-english_2022}. The model can be run interactively on the page and deployed to the cloud by all visitors to the website since the model is open source. The left column provides documentation on the risks, limitations, and biases of the model, but no formal and comprehensive evaluation of the identified bias properties are provided in the dataset evaluation listing of the lower right. Understanding the biases of the model is left as an exercise to the developer making use of the model. Instead, flat performance properties like accuracy and F1 scores are presented and verified by Hugging Face staff. Given the model has been downloaded more than 6 million times, it is likely that the vast majority of model deployers have not engaged in any sort of formal governance program. Note: all the contents of the page are found on the Hugging Face website, but we have deleted some contents so all the elements in the screen capture will be rendered together.}
\label{fig:huggingface}
\end{figure}

The model sharing and deployment company Hugging Face in one of their language models (see Figure \ref{fig:huggingface}) indicates the model presents significant biases but does not formally evaluate those biases for the community. Instead, they provide a series of top level performance properties. Model monitoring companies close the gap between data evaluation and human oversight by incorporating visual analytic user interfaces into the data logging functionality. These include Neptune.ai, Arize, WhyLabs, Grafana+Prometheus, Evidently, Qualdo, Fiddler, Amazon Sagemaker, Censius, ArthurAI, New Relic, Aporia, TruEra, Gantry, and likely others in this quickly expanding market space (see: \cite{czakon_best_2021} for a rundown).

These systems are essentially data science platforms -- they support a person exploring data as it is streaming in. What they don't do without additional effort is codify requirements in such a way that they can be checked automatically and continuously. While it is possible to continually staff a data science project with personnel applying governance requirements, the value of data-centric governance is the formalization of the monitoring activity so that people do not continuously need to watch the data as it flows in.


\textbf{Continuous Integration/Continuous Delivery (CI/CD).}

The final system of continuous assurance is one that wraps the governance program in software systems that continuously check for compliance with requirements. Should those requirements be violated, then the system can either automatically move to a fail safe mode (typically this means shutting down) or alert humans to begin evaluating the system for potential safety and fairness issues.

\begin{figure}[ht]
\centering
\includegraphics[width=0.7\textwidth]{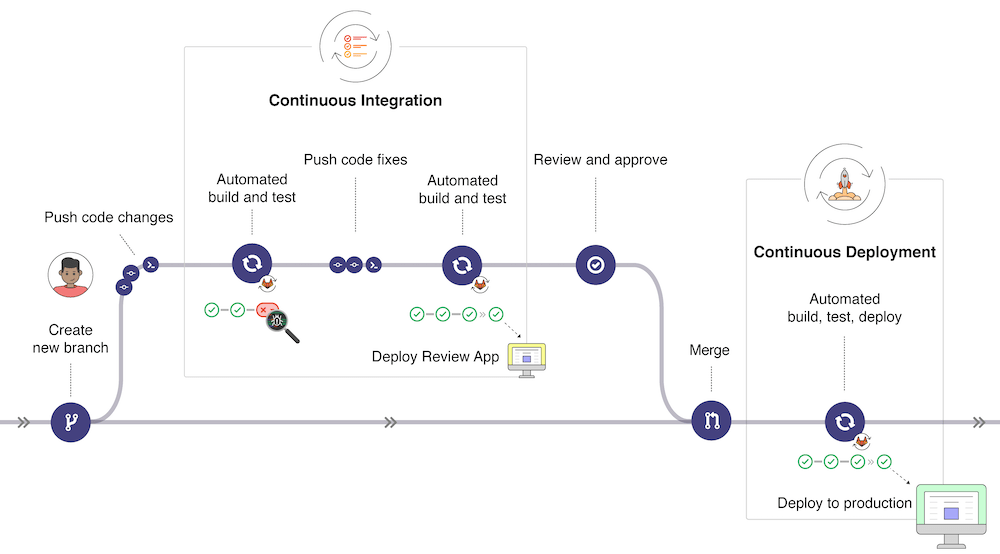}
\caption{Continuous integration and delivery processes as supported by the DevOps company GitLab. The process begins with a developer creating a new version of the code, which when pushed to the GitLab server triggers a cycle of automated testing and deploying fixes to any failing tests. When the tests pass and are approved, they then move to the delivery process where additional tests are run by GitLab before the system automatically deploys to the world as the ``production'' version of the system. From \cite{gitlab_cicd_2022}}
\label{fig:gitlab}
\end{figure}

\begin{figure}[ht]
\centering
\includegraphics[width=0.7\textwidth]{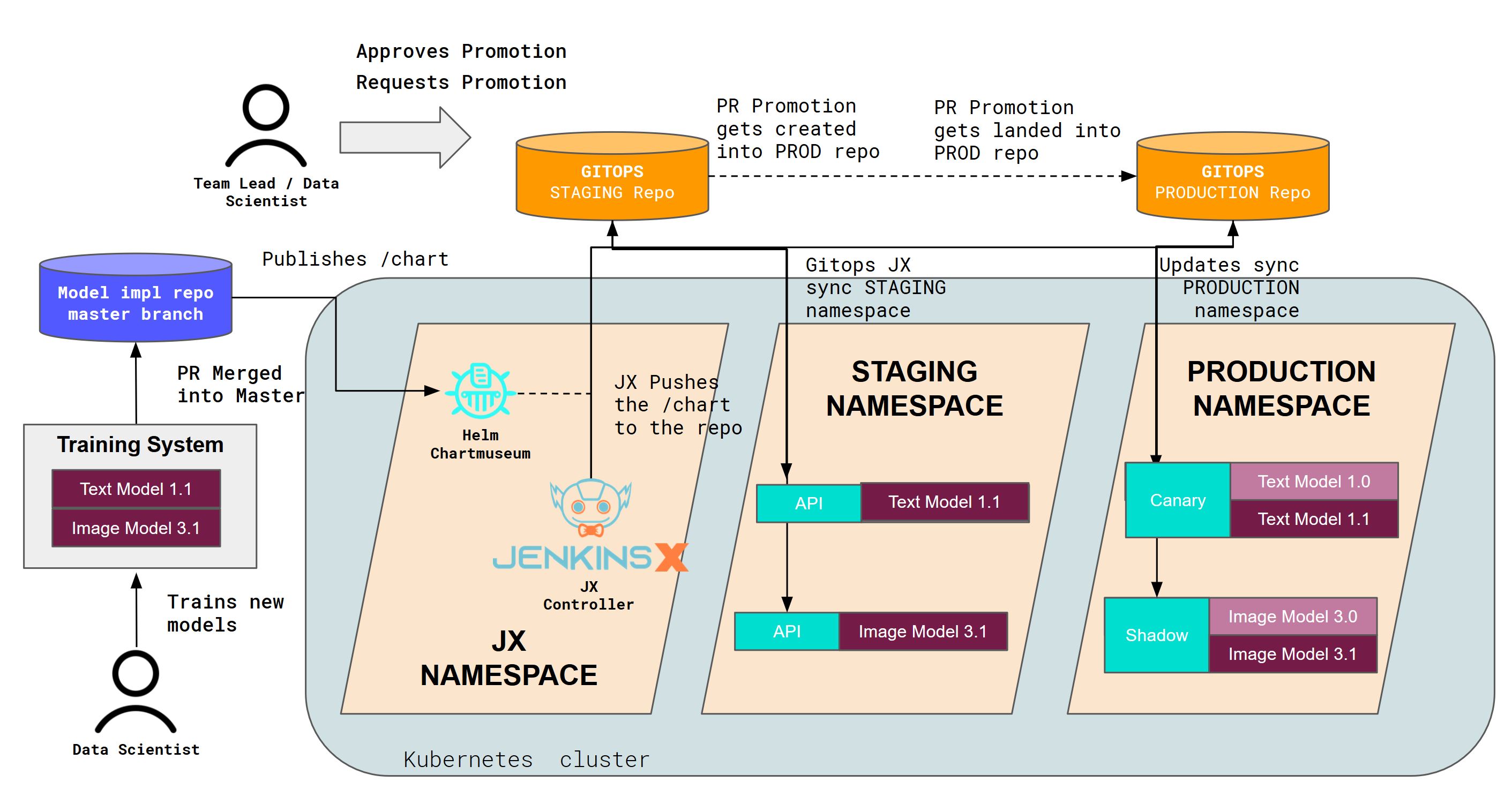}
\caption{Seldon Core's model-centric view of continuous integration and delivery. Like GitLab in Figure \ref{fig:gitlab}, the process begins with a developer (here a data scientist) updating the implementation, which then goes into a scalable computing environment known as a ``Kubernetes cluster.'' The scalable computing environment then runs tests and an approver can decide whether the update goes to a staging environment for further testing and/or the live production environment. As a data-centered practice, the system has multiple versions deployed simultaneously so each can be statistically compared in their live performance. \cite{seldon_core_seldon_2022}}
\label{fig:seldon}
\end{figure}

Most software today is developed with systems for continuous integration (i.e., systems that continuously test for new failures or ``regressions'') and continuous delivery (i.e., systems for the deployment of a model into the real world). For instance, the developer operations (DevOps) platform GitLab provides the ability to integrate, test, and deploy software updates as shown in Figure \ref{fig:gitlab}. Seldon Core similarly provides systems shown in Figure \ref{fig:seldon} supporting humans making the decision of whether a model should be deployed after reviewing the system performance as reported in testing.

\section{Conclusion}
\label{sec:conclusion}
The data needs, systems, and processes we have introduced may seem like a large burden, but they should be viewed in light of the benefits they provide. AI systems can operate at unlimited scale and speed. With strong data-centric governance, the quality of those solutions improves, fewer AI efforts fail, and system have a much longer useful life. Data-centric governance explicitly accounts for the hidden costs of program failures and moves uncertainties into reliable process steps.

When enacting a data-centric governance approach, the first step is to constitute or contract with teams capable of carrying out each of the functions we identified. With appropriate teams in place, it is possible to capture the insights and requirements of human auditors, stakeholders, and other governance process participants in a way that will most benefit the deployment of the system -- rather than block deployment at the 11th hour.

\highlight{
\textbf{Is it supposed to be this hard?} 

As a final case study, we appeal to the history of oceanic shipping, steam boilers, and electricity. Each were extremely risky in their early histories and regularly lead to loss of life and steep financial losses. Today shipping is very safe, steam boilers don't regularly explode, and electricity is in every modern home with little risk of electrocution or fire. The story of all these industries becoming as safe as they are today is the story of the insurance industry. Insurance companies assess risks and charge fees according to those risks. When something is more expensive to insure, then you know it is also riskier than its competitors. Thus companies have an incentive to sail calm waters, design safer boilers, and standardize electrical wiring.

With a track record of anticipating emerging risks (e.g. for insuring the performance of green technologies), multinational insurance company, MunichRe, began offering insurance for AI systems \cite{munich_re_insure_2022}. Scoped around insuring the performance of AI products (e.g., how well a system filters online content for moderation), the ``aiSure'' product requires the development of a suite of tools for monitoring system performance. In effect, MunichRe has arrived at a similar conclusion to that of data-centric governance -- the operating conditions must be defined and continuously assessed. When deploying an AI system to the world, if you do not believe that MunichRe would be able to insure the system's performance, then it is not functionally governed.

\takeaway{Is it supposed to be this hard? Yes! But it is worth it.}}

With systems of continuous assurance built into a solution from the start, governance becomes a product asset rather than a liability. We can build a more equitable and safer future together with AI.

\begin{ack}
We gratefully acknowledge the review and contributions of Andrea Brennen and Jill Crisman in the production of this work. As a position paper from and for the communities of test and evaluation, verification and validation, AI safety, machine learning, assurance systems, risk, and more, this paper would not be what it is without broad and varied input. We invite your review and feedback to improve the concepts and their communication to varied audiences.

\textbf{Funding.} This work was made possible by the funding of IQT Labs.
\end{ack}


\bibliography{zotero-sean}


\end{document}